\documentclass[showpacs,twocolumn,prb]{revtex4-1}
\usepackage{graphicx}
\graphicspath{{figures/}} % Location of the graphics files
\usepackage{amsmath}
\usepackage{enumitem}

\newcommand{\bwt}{\begin{widetext}}
\newcommand{\ewt}{\end{widetext}}

\newcommand{\be}{\begin{equation}}
\newcommand{\ee}{\end{equation}}
\def\bea {\begin{eqnarray}}
\def\eea {\end{eqnarray}}

\usepackage{hyperref}
\hypersetup{
    colorlinks=true,
    linkcolor=blue,
    filecolor=magenta,      
    urlcolor=blue,
}

\usepackage{color} 
\newcommand{\itt}{\it}

\usepackage{comment}
\def\comment#1{}

\usepackage{ulem}   % to strike things out
\begin{document}

\title{Impact of Retardation in the Holstein-Hubbard Model: a two-site Calculation}

\author{F. Marsiglio}
\affiliation{Department of Physics, University of Alberta, Edmonton, AB, Canada T6G~2E1}

\begin{abstract}
Eliashberg theory provides a theoretical framework for understanding the phenomenon of superconductivity
when pairing between two electrons is mediated by phonons, and retardation effects are fully accounted for.
However, when a direct Coulomb interaction between two electrons is also present, this interaction is only
partially accounted for. In this work we use a well-defined Hamiltonian, the Hubbard-Holstein model, to examine
this competition more rigorously, using exact diagonalization on a two-site system. We find that the direct electron-electron
repulsion between two electrons has a significantly more harmful effect on pairing than suggested through the standard
treatment of this interaction.
\end{abstract}

\date{\today }
\maketitle

\section{introduction}

Migdal-Eliashberg (ME) theory\cite{migdal58,eliashberg60a,eliashberg60b,nambu60,scalapino69,allen82,rainer86,carbotte90,marsiglio08,marsiglio20}
represents the state-of-the-art methodology for computing superconducting properties of various so-called electron-phonon, or 
conventional\cite{physicac2015} superconductors. The theory builds on the pairing formalism of the Bardeen-Cooper-Schrieffer (BCS) theory of
superconductivity\cite{bardeen57} and shares a number of common traits with BCS, but also differs in some significant aspects.

Both formalisms are mean field theories, although BCS is also a variational wave function at zero temperature. They are both based on electron
pairing; however BCS theory is based on a pairing wave function, while ME theory is based on an anomalous pairing Green function.\cite{gorkov58}
BCS theory relies on an underlying normal state Fermi liquid; ME also does this, but a further justification is provided by the Migdal
approximation,\cite{migdal58} which is often cited as a ``theorem''.  It turns out that electron-phonon interactions in the normal state can and do lead
to polaron and bipolaron effects,\cite{alexandrov01} whereby the electrons acquire large effective masses. Alexandrov and coworkers\cite{alexandrov01}
repeatedly argued since the 1980's that polaron effects would overwhelm the Migdal Fermi Liquid \textit{and} nonetheless lead to superconductivity. 
His was a lone voice in the wilderness.\cite{remark_alexandrov_ranninger}

Early Quantum Monte Carlo (QMC) studies of a simple local model for the electron-phonon interaction, the so-called Holstein model,\cite{holstein59}
in one\cite{hirsch82} and two\cite{scalettar89,marsiglio90,marsiglio91,noack91,vekic92} dimensions demonstrated a propensity towards superconductivity,
and even found quantitative agreement with ME calculations, provided the renormalized ME theory was used. Here, ``renormalized'' means that
a phonon self energy is included in the calculation; this was not the practice in the preceding three decades of calculation since the phonon spectrum
required in ME was usually taken from tunnelling experiments,\cite{mcmillan69} and therefore already contained renormalization effects.

While the positive aspects of the QMC calculations were generally emphasized in these papers, little was said about the fact that 
\begin{enumerate}[label=(\alph*)]
\item generally 
only very weak coupling strengths were reported ($\lambda \approx 0.25$, where $\lambda$ will be defined below and roughly 
corresponds to the dimensionless mass enhancement parameter).
\item the phonon energy ($\hbar \omega_E$) used was always of order the hopping parameter, $t$, 
rather than the more physical regime, $\hbar \omega_E << t$, and 
\item the 
competing effect of the Coulomb repulsion, represented for this local model by the Hubbard $U$,\cite{hubbard63} was typically ignored.
\end{enumerate}

These choices could 
have various reasoning behind them; from our point of view (a) was necessary to get reasonable results that had a chance of agreeing with 
ME (this therefore constituted ``confirmation bias'' for the present author), 
(b) was required so the sampling algorithm would remain ergodic; having two very different energy scales for the electrons and the phonons would
lead to very different equilibration times in the simulation of these two degrees of freedom, and therefore made it very difficult to attain accurate
results, and (c) we did look into the effect of the Hubbard $U$, and it immediately
squashed all hope of superconductivity. For this latter point, the present author rationalized that this occurred because of the relatively high phonon frequency we
were forced to adopt [because of point (b)], and therefore we left for another day the demonstration of the so-called pseudopotential 
effect.\cite{bogoliubov59,morel62} 

The pseudopotential effect results in an effective Coulomb potential that is much smaller than the bare Coulomb repulsion because
electron-electron correlations induced by the electron-phonon interactions keep two electrons apart at the same time. This retardation effect is
at the heart of ME theory, but is also minimally contained in BCS theory through the presence of a cutoff in the interaction in wave vector space. To our
knowledge nobody at the time of these early QMC studies reported the inability to see this pseudopotential effect in their numerical work.

Since these early QMC studies, a number of follow-up studies on the Holstein-Hubbard model
have been published. Niyaz et al.\cite{niyaz93} focused on the charge-density-wave (CDW)
instability at half-filling, and von der Linden and coworkers\cite{berger95}  used a projector-QMC technique to benchmark self-consistent Green function
calculations that include the effect of the Coulomb repulsion, $U$. Hohenadler and coworkers also developed a QMC method\cite{hohenadler04,hohenadler05}
based on the Lang-Firsov transformation, and were able to more accurately explore regimes ($\omega_E < t$) that were previously inaccessible.

More recently there has been a resurgence in QMC studies\cite{nowadnick12,johnston13,ohgoe17,esterlis18,weber18,bradley21}
and in self-consistent Migdal-Eliashberg calculations\cite{dee19,schrodi21}
of the Holstein model; moreover, several of these references have incorporated the Hubbard interaction as competition. 
A summary of some of the Monte Carlo work is provided in Ref.~\onlinecite{chubukov20}; the conclusion, based on work over the past three decades is
that, while a ``breakdown'' of Migdal-Eliashberg theory clearly occurs for fairly weak electron-phonon couplings, a hope remains that Migdal-Eliashberg
theory can still describe superconductivity. Part of this argument is by analogy (see the Discussion section in Ref.~\onlinecite{chubukov20}), and partly
it is because there remain effective couplings (called $\lambda_0^{\rm eff}$ and $\lambda^{\rm eff}$ in 
Refs.~[\onlinecite{marsiglio90},\onlinecite{marsiglio91}] and
Ref.~[\onlinecite{chubukov20}], respectively) that can exceed the bare coupling considerably. However, this requires considerable phonon softening
over a wide temperature range, and this has not been properly subjected to testing in materials like Pb and Nb.\cite{stedman67,shapiro75,aynajian08}

In addition a number of DMFT studies\cite{bauer11,bauer12,bauer13} have specifically addressed the role of the Coulomb repulsion; we will return to this
reference later, as the present paper will address the role of Coulomb interactions as well, but by using a simple two-site model
for the Holstein-Hubbard model, and performing an exact diagonalization study.

Exact diagonalization studies for the Holstein-Hubbard model began with Ranninger and Thibblin.\cite{ranninger92} Subsequently
Ref.~[\onlinecite{marsiglio95}]
provided an exact demonstration (see Fig.~7 of that paper) of the pseudopotential effect previously determined through ``back-of-the-envelope" type
calculations.\cite{bogoliubov59,morel62} It is the purpose of this paper to quantify the strength of the pseudopotential effect. Because they are
so straightforward, we will utilize numerical solutions (as in Refs.~[\onlinecite{marsiglio93,alexandrov94,marsiglio95}]), and bypass the more elegant analytical
solutions (for the two-site problem alone) given in Refs.~[\onlinecite{rongsheng02},\onlinecite{berciu07}]. We also make note of the very powerful solution for one
or two electrons provided in Refs.~[\onlinecite{bonca99},\onlinecite{bonca00}] which allows for an exact numerical solution for the single polaron and bipolaron
problems on an infinite lattice; as noted previously,\cite{berciu07} the two-site solutions seem to capture many aspects of the infinite lattice solution.

The outline of this paper is as follows. We define the Hamiltonian in the next section. Diagonalization can be performed for any number of electrons
($N_e = 1,2, 3$ or $4$)
and any number of phonons on each of the two sites, subject to some cutoff. However, more efficient results are obtained by transforming both the
phonon and electron operators\cite{ranninger92} as demonstrated in the Appendix. We then present results for various electron-phonon coupling values 
and phonon frequencies, and demonstrate the devastating effect of the Hubbard $U$. We make the connection of these two-site results with those
for an infinite system,\cite{bonca00} and then draw conclusions.

\section{The Holstein-Hubbard model}

\subsection{The Hamiltonian}

The Holstein-Hubbard model is written as a sum of different contributions in real-space,
\begin{equation}
H = H_{\rm el} + H_{\rm ph} + H_{\rm e-ph} + H_{\rm el-el},
\label{hh_ham1}
\end{equation}
where
\begin{equation}
H_{\rm el} = \sum_{\langle ij \rangle, \sigma} \left[ t_{ij} c^{\dagger}_{i\sigma} c_{j\sigma} \ + \  h.c. \right]
\label{ham_el}
\end{equation}
represents the electron-hopping term for an electron with spin $\sigma$ from site $i$ to site $j$ with amplitude $t_{ij}$ and vice-versa.
Electron creation (annihilation) operators with spin $\sigma$ at site $i$ are represented by $c^{\dagger}_{j\sigma}$ ($c_{j\sigma}$), and
the electron number operator is given by  $n_{j\sigma} = c^{\dagger}_{j\sigma} c_{j\sigma}$. Typically,
nearest-neighbour hopping only is included, so $t_{ij} = -t$ for $i,j$ nearest neighbours with lattice spacing $\ell$. Often periodic boundary conditions are used to
speed convergence to the thermodynamic limit; in this work, even though we utilize just two sites, we will still use periodic boundary conditions, as we find
these solutions allow us to match various parameters with the known infinite solutions.

All other terms are diagonal in real space. This lattice model assumes that an atom occupies each site, and is in its equilibrium position, except for small
vibrations about each position, represented by the operator $x_j$. In a real solid, these displacements are connected with one another, but in the
Holstein model it is assumed that the vibrations are completely local. Hence the phonon Hamiltonian is given by
\begin{equation}
H_{\rm ph} = \sum_{j} \left[ {P_j^2 \over 2M} + {1 \over 2} M \omega_E^2 x_j^2 \right],
\label{ham_ph}
\end{equation}
where $P_j$ is the momentum operator for the ion of mass $M$ at site $j$; this operator is the conjugate variable to $x_j$. We have also
introduced the Einstein oscillator frequency, $\omega_E$ related to the spring constant $k$ by $k \equiv M\omega_E^2$.
Because this model
is local we can introduce Dirac raising ($a_j^\dagger$) and lowering ($a_j$) operators in real space,
\begin{equation}
x_j \equiv \sqrt{\hbar \over 2M\omega_E} \left(a_j^\dagger + a_j\right) \ \ \ P_j \equiv i\sqrt{\hbar M \omega_E \over 2} \left(a_j^\dagger - a_j\right),
\label{dirac}
\end{equation}
in terms of which (leaving out the vacuum energy) the phonon part of the Hamiltonian is simply
\begin{equation}
H_{\rm ph} = \hbar \omega_E \sum_{j} a_j^\dagger a_j.
\label{ham_ph_b}
\end{equation}

The electron-phonon coupling is taken to linear order in the displacement, resulting in the minimal model,
\begin{equation}
H_{\rm e-ph} = -\alpha \sum_{j} (n_{j\uparrow} + n_{j\downarrow}) x_j.
\label{ham_ph_a}
\end{equation}
In terms of the phonon raising and lowering operators, this can be written as
\begin{equation}
H_{\rm e-ph} = -g\omega_E \sum_{j} (n_{j\uparrow} + n_{j\downarrow}) (a_j^\dagger + a_j),
\label{ham_ph_b}
\end{equation}
where a new (dimensionless) coupling constant $g$ is introduced in terms of the original coupling constant $\alpha$:
\begin{equation}
g\omega_E \equiv \alpha \sqrt{\hbar \over 2 M \omega_E}.
\label{e-ph_coupling}
\end{equation}
In fact in the superconducting literature, a dimensionless coupling constant $\lambda$ is generally used. Here it is defined by
\begin{equation}
\lambda \equiv {1 \over W} {\alpha^2 \over M \omega_E^2} \equiv {1 \over W} {2g^2 \hbar \omega_E},
\label{lambda}
\end{equation}
where $W$ is the electronic bandwidth and $1/W$ represents an average electronic density of states. Since we are using two sites with
periodic boundary conditions, we use $W = 4t$.

Finally the electron-electron repulsion is described by the simple Hubbard model,
\begin{equation}
H_{\rm el-el} = U \sum_{j} n_{j\uparrow} n_{j\downarrow},
\label{ham_ee}
\end{equation}
with relevant energy scale $U$, representing the on-site Coulomb repulsion for two electrons in the same orbital.

Further simplifications specifically for the two-site Hamiltonian are found in the Appendix. For example, total momentum is conserved, and
the Hilbert space can be divided into sectors with different total momentum; for two sites, these are $q_{\rm tot} = 0$ and $q_{\rm tot} = \pi/\ell$.

% fig. 1 2site
\begin{figure}[tp]
\begin{center}
\includegraphics[height=3.4in,width=2.4in,angle=-90]{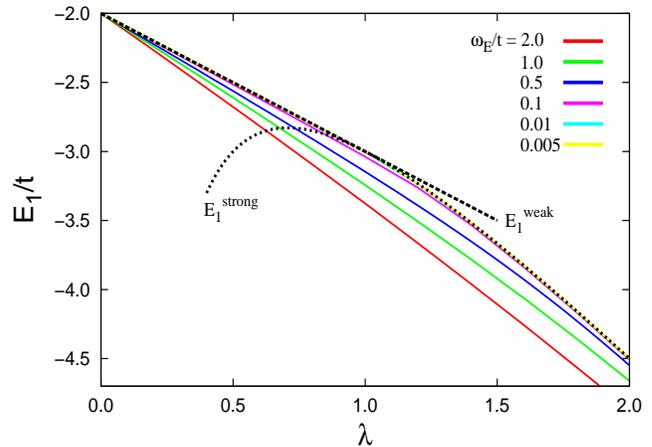}
\end{center}
\caption{Single particle ground state energy, with total wave vector $q_{\rm tot} = 0$ vs. 
$\lambda \equiv 2g^2 \hbar \omega_E/W$ for a variety of values of $\hbar \omega_E$, starting from $\hbar \omega_E = 2.0t$ 
(lower-most (red) curve at $\lambda = 2$) and progressing upwards as $\hbar \omega_E$ decreases. Results are shown 
for $\hbar \omega_E = 2t$, $1t$, $0.5t$, $0.1t$, $0.01t$, and $0.005t$. The curves corresponding to the latter three 
values of $\hbar \omega_E$ are barely distinguishable from one another, showing that the adiabatic
limit has been achieved. Also shown are weak and strong coupling results \cite{marsiglio95} with dashed and dotted curves, 
respectively. The weak coupling result (with $\omega_E \rightarrow 0$) is $E_{1}^{\rm weak} = -2t - t\lambda$, and
the strong coupling result is $E_{1}^{\rm strong} = -2t \lambda - t/\lambda$. As discussed in Refs.~[\onlinecite{kabanov93,li10,li12}] a weak coupling
regime does not actually exist in 1D (it does in 2D and 3D) and exists here only because we use two sites. Note the very good accuracy of the strong
coupling approximation for $\lambda {{ \atop >} \atop {\approx \atop }} 1$ for small phonon frequencies --- it covers the exact numerical results for the three
lowest phonon frequencies for this range of coupling.}
\label{fig1_2site}
\end{figure}

\subsection{One electron}

Following the procedure outlined in the Appendix, the eigenstates and eigenvalues are calculated in the one-electron sector. Figure~\ref{fig1_2site} shows the ground state energy (always in the $q_{\rm tot} = 0$ subspace) as a function of the electron-phonon coupling
strength, $\lambda$, for a variety of phonon frequencies. Also shown are the adiabatic approximations for two sites from 
Ref.~\onlinecite{marsiglio95}. Note that the strong coupling approximation differs slightly for two sites from the result obtained with open boundary conditions.

Also note that to properly converge the results near $\lambda = 2$
and for small phonon frequency requires $\approx 2000$ phonons. While we cannot calculate the effective mass, for example, with these two-site calculations,
we know from many other studies now that at the very least the regime $\lambda > 1$ should be excluded from further consideration, as it results in highly polaronic single particle states.\cite{remark1} As discussed further below, in the present study this breakdown is signalled by an abundance of phonons in the ground state.

\subsection{Two electrons}

\begin{figure}[tp]
\begin{center}
\includegraphics[height=3.4in,width=2.4in,angle=-90]{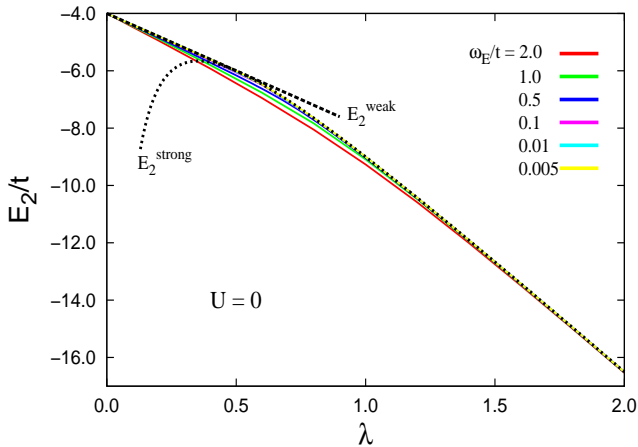}
\end{center}
\caption{Two-particle ground state energy with total wave vector $q_{\rm tot} = 0$ vs. $\lambda \equiv 2g^2 \hbar \omega_E/W$ for a variety of
values of $\hbar \omega_E$, starting from $\hbar \omega_E = 2.0t$ (lower-most red curve) and progressing upwards as
$\hbar \omega_E$ decreases. Results are shown for $\hbar \omega_E = 2t$, $1t$, $0.5t$, $0.1t$, $0.01t$, and $0.005t$. For 
$\lambda {{ \atop >} \atop {\approx \atop }} 1$ the curves for the lowest four frequencies are barely distinguishable from one another. Also shown are weak and strong
coupling results \cite{marsiglio95} with dashed and dotted curves, respectively. The weak coupling result for low phonon frequency is 
$E_{2}^{\rm weak} = -4t(1+\lambda)$, and the strong coupling result is $E_{2}^{\rm strong} \approx -8\lambda t - t/\lambda$.
The strong coupling result is very accurate for $\lambda {{ \atop >} \atop {\approx \atop }} 0.8$, and essentially covers the numerical
results for all phonon frequencies. $U=0$ in all cases.
}
\label{fig2_2site}
\end{figure}

Following the Appendix, we now revert to the two-particle subspace.
Figure~\ref{fig2_2site} shows the ground state energy (always in the $q_{\rm tot} = 0$ subspace) as a function of the electron-phonon coupling
strength, $\lambda$, for a variety of phonon frequencies. 
Note the results converge to the strong coupling result for all frequencies, $E_{2}^{\rm strong} \approx -8\lambda t - t/\lambda$, and the weak coupling
limits for low phonon frequency, $E_{2}^{\rm weak} \approx -4t - 4\lambda t$ ($\omega_E << t$) as indicated. For high phonon frequency,
$E_{2}^{\rm weak} \approx -4t - 6\lambda t$ ($\omega_E >> t$) (not shown).\cite{marsiglio95} As was the case in the one-electron sector, these results 
differ slightly from the result for two sites with open boundary conditions.

\begin{figure}[tp]
\begin{center}
\includegraphics[height=3.4in,width=2.4in,angle=-90]{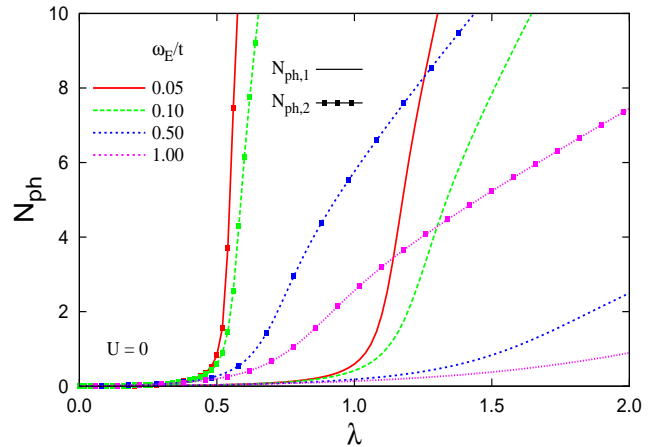}
\end{center}
\caption{The total number of phonons, corresponding to the difference coordinate ($d_\pi^\dagger$ and $d_\pi$ in the Appendix) in the ground state
wave functions for one electron (curves) and for two electrons (curves with square symbols) vs. $\lambda$. Results are shown for various
values of $\hbar \omega_E$ as indicated, and for $U=0$. As the phonon frequency decreases more phonons are present in the ground state, particularly for
two electrons. When the average number of phonons exceeds $N_{ph} \approx 5$ the state is very polaronic.
}
\label{fig3_2site}
\end{figure}

Because we have restricted our calculations to two sites, the effective mass is not readily accessible. However, a proxy for the effective mass
is the number of phonons in the ground state --- both this quantity and the effective mass will increase 
rapidly as polaronic effects become dominant. In Fig.~\ref{fig3_2site} we show
the number of phonons in the ground state for both the one-electron and two-electron sectors. Note that for phonon frequency of order the
hopping parameter the number of phonons is reasonably low as a function of the coupling strength. However, for more realistic phonon frequencies,
$\omega_E << t$, the number of phonons, particularly in the two-particle calculation, grows rapidly with increasing coupling strength. In practice,
for superconducting materials where polaron effects are {\it not observed}, this observation constrains the range of reasonable coupling strengths.

% fig. 2 2site

% fig. 3 2site

\section{The Binding Energy}

The binding energy $\Delta$ for two electrons is given by the simple relation
\begin{equation}
\Delta \equiv 2E_1 - E_2,
\label{binding}
\end{equation}
where $E_1$ and $E_2$ are the single- and two-particle ground state energies as calculated above. A negative result for $\Delta$ indicates
no binding. The significance of the binding of two electrons on a two-site lattice has been discussed previously,\cite{berciu07} and in particular
a careful delineation of on-site (S0) and neighbouring site (S1) type pairing was made.\cite{proville98,bonca00} Here we wish to emphasize the extent to which any binding
persists in the presence of an on-site Coulomb interaction. The basic idea dating back to Refs.~[\onlinecite{bogoliubov59},\onlinecite{morel62}] (see also
Ref.~[\onlinecite{marsiglio89}]) is that the Coulomb interaction is effectively reduced due to the much slower pairing induced by the electron-phonon
interaction. The idea is that the bare interaction $U$ will be reduced to $U^\ast (\omega_E)$, where \cite{bogoliubov59}
\begin{equation}
U^\ast(\omega_E) = {U \over 1 + {U \over W} \ln{\left({W \over 2\hbar \omega_E}\right)}},
\label{pseudo}
\end{equation}
where $W=4t$ is the electronic bandwidth. One of the important consequences about this approximate formula is that a significant reduction occurs, even
as $U \rightarrow \infty$. This limit has undoubtedly contributed to the widespread adoption of a quasi-universal value for this value,
$\mu^\ast(\omega_E) \equiv U^\ast(\omega_E)/W~\approx~0.1$ in dealing with superconductors; moreover, more recently researchers have generally neglected that this value 
has a dependence on the reduced frequency scale, $\omega_E$, and so for much larger $\omega_E$ (as expected in systems involving hydrogen
vibrations, for example), the reduction from $U$ should be significantly less.

\begin{figure}[tp]
\begin{center}
\includegraphics[height=3.0in,width=2.2in,angle=-90]{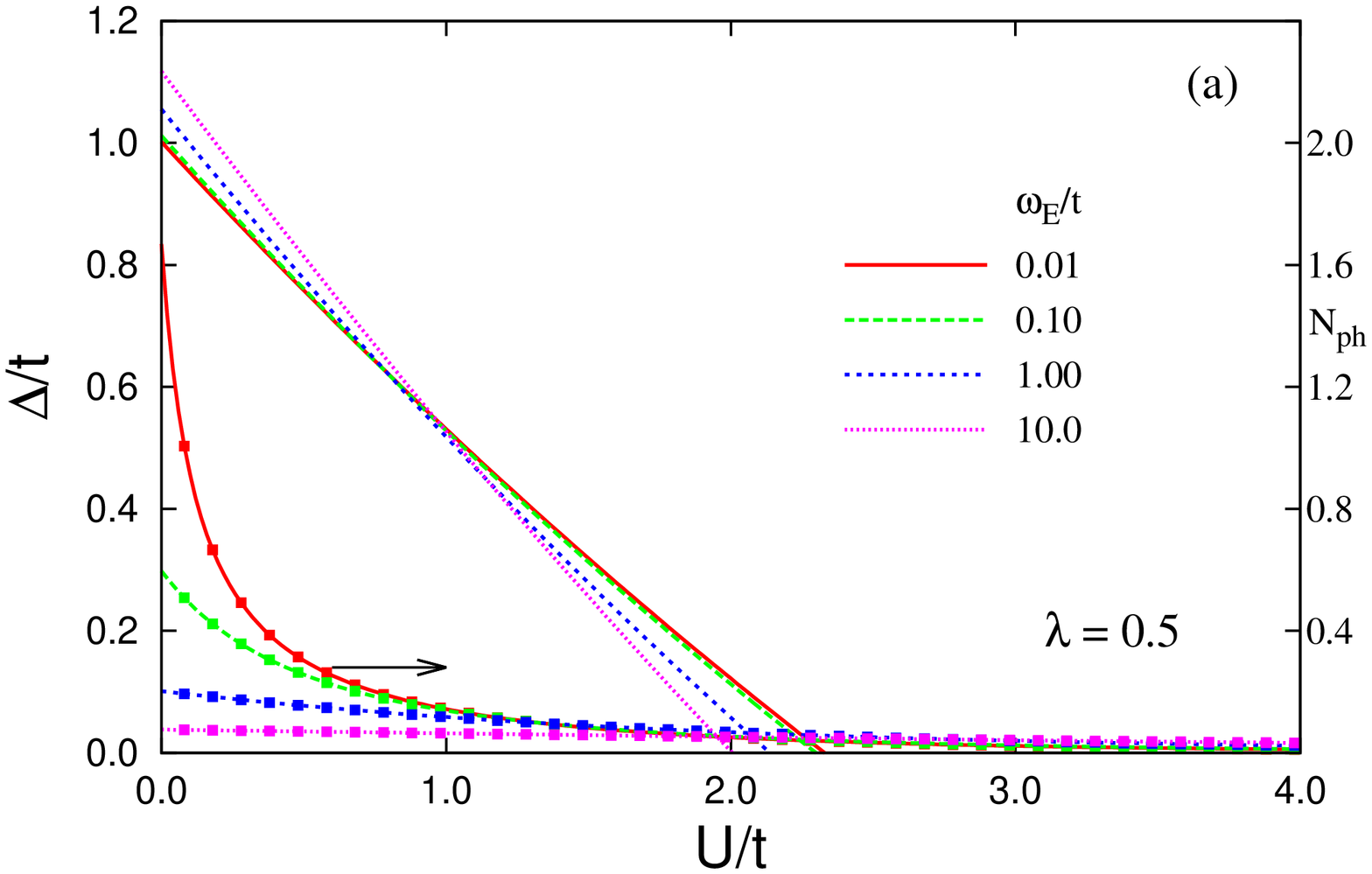}
\includegraphics[height=3.0in,width=2.2in,angle=-90]{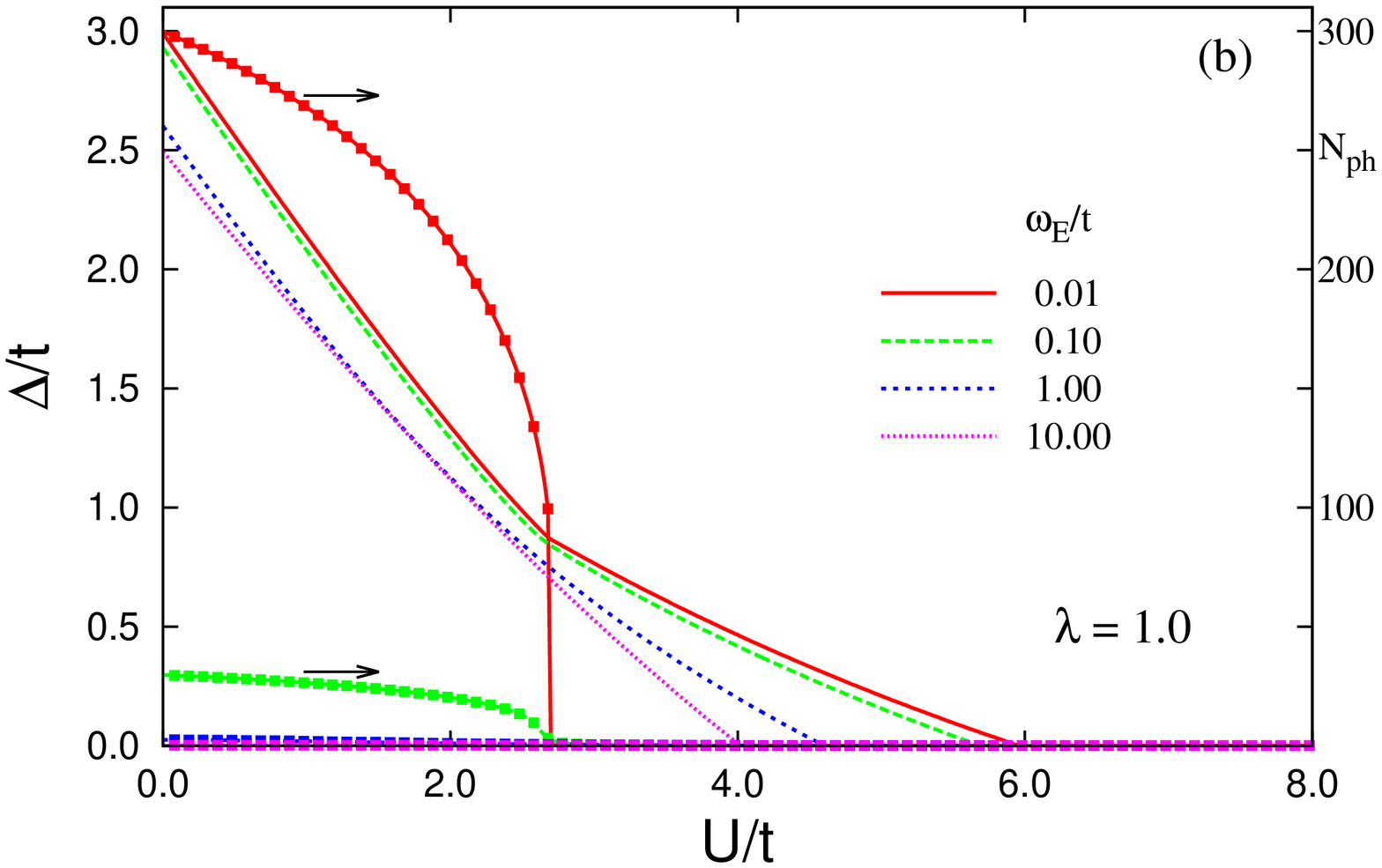}
\end{center}
\caption{The binding energy $\Delta/t$ vs the on-site Coulomb repulsion $U/t$ for various values of $\omega_E$ as indicated and for (a) $\lambda = 0.5$ and
(b) $\lambda = 1$. Retardation ($\hbar \omega_E <<t$) clearly allows binding even when $U > W\lambda$, as is especially evident in (b). Also shown 
(right side scale) are the number of phonons
present in the two-particle ground state (curves with symbols); these best indicate the crossover from $S0$ to $S1$ type bipolaron ground state. Note that in both
cases, binding ceases at some point below the $U = 8\lambda t$ estimate.\cite{bonca00}
}
\label{fig4_2site}
\end{figure}

In Fig.~\ref{fig4_2site} we show the binding energy $\Delta$ vs. $U/t$ (curves without symbols) for various phonon frequencies for (a) $\lambda = 0.5$ and
(b) $\lambda = 1.0$. Also shown (curves with symbols) are average number of phonons in the two-particle ground state (right scale) so that one can get 
a feel for the degree of polaronic effects. Figure~\ref{fig4_2site}(a) corresponds to the weak coupling regime, while Fig.~\ref{fig4_2site}(b) exemplifies
the more strongly coupled regime. In each case, four different curves are shown, corresponding to different values of $\omega_E/t$ as
described in the caption. As expected, the binding energy goes to zero (no binding) for sufficiently large $U/t$. Clearly binding is present
at $U=0$, but actual values of $U$ in such a model are expected to exceed $4\lambda t$ for stability reasons, i.e. $U/t \ge 2$ in (a) and $U/t \ge 4$ in (b).
Binding ceases to occur for sufficiently large values of $U$ in either case.

Regardless of the legitimacy of the magnitude of $U$, the figure does make clear the effect of retardation. While no binding is present in
Fig.~\ref{fig4_2site}(a) for $U>2t$, and in (b) for $U>4t$ for a very (unrealistically) high phonon
frequency, $\omega_E = 10t$, it is clearly present for lower phonon frequencies, as illustrated by the result for $\omega_E = 0.01t$, particularly in (b). Also note
that the binding changes its character, as has been well described in Refs.~[\onlinecite{bonca00},\onlinecite{berciu07}]. In particular, focussing on (b), note
the kink in the binding energy curve (for $\omega_E = 0.01t$), accompanied by the precipitous change in the two-particle phonon occupation 
at $U/t \approx 2.7$. The bound pair below this value is primarily on-site ($S0$), while above this value of $U$ it is primarily a nearest neighbor pair ($S1$). These
are designated as $S0$ and $S1$ bipolarons, respectively.\cite{proville98,bonca00,berciu07} The $S0$ bipolaron has a very unrealistically large
effective mass; here this property is manifested in the large number of phonons in the ground state, which rapidly becomes small as the bipolaron transitions
to the $S1$ type.

% fig. 4 2site

There is a prolonged region of non-zero binding beyond this $S0-S1$ crossover point, {\it provided the
phonon frequency is sufficiently low to allow retardation effects}. An additional complication with increasing values of $\lambda$ is that the 
single-particle phonon occupation becomes very large, indicative of a very polaronic material. Therefore unless these polaronic effects are 
present in the material in the normal state,
this model would be ill-suited to describe such a material. Moreover, if this were the case, then the superconductivity would be unconventional, and have
attributes better described by Alexandrov's prescription\cite{alexandrov01} than by Migdal-Eliashberg theory.

% fig. 5 2site
\begin{figure}[tp]
\begin{center}
\includegraphics[height=3.0in,width=2.2in,angle=-90]{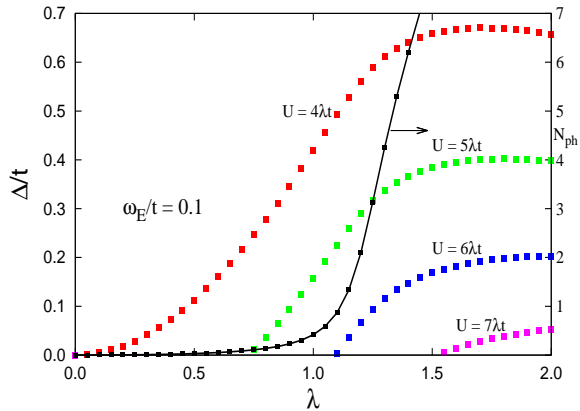}
\end{center}
\caption{The binding energy $\Delta/t$ vs $\lambda$, with $U$ also increasing as $\lambda$ increases, i.e. $U = x\lambda t$, and $x = 4, 5, 6$ and $7$ as indicated.
Note that the binding peaks as a function of $\lambda$. Also shown (right side scale) is the number of phonons in the one-particle
ground state, which exceeds 4 for $\lambda {{ \atop >} \atop {\approx \atop }} 1.3$, indicating large polaronic effects.\cite{remark1}
}
\label{fig5_2site}
\end{figure}

As already mentioned, for reasons of lattice stability, the minimum expected value of $U$ is $W\lambda$ where $W = 4t$ is the one-dimensional
electronic bandwidth.\cite{cohen72} 
In Fig.~\ref{fig5_2site} we show the binding energy as a function of $\lambda$, with $U = 4\lambda t$, $5\lambda t$, $6\lambda t$, 
and $7\lambda t$ as indicated. This figure uses $\hbar \omega_E = 0.1t$, which is sufficiently small that one can take full advantage of retardation to overcome
the direct Coulomb repulsion represented by $U$.
The result for $U=8\lambda t$ is zero as expected from Fig.~\ref{fig4_2site} and as determined already in Refs.~[\onlinecite{bonca00},\onlinecite{berciu07}]. We also
display the number of phonons in the ground state for the single-electron sector (curve with symbols, using right-side scale). Note that already
at $\lambda \approx 1.3$ the number of phonons present exceeds 4, which indicates the ground state as a highly polaronic character with a very
heavy effective mass.\cite{remark1}

\section{Discussion}

As pointed out already by Berciu,\cite{berciu07} comparison with calculations for infinite systems indicates that this two-site calculation is relevant
for bulk systems. There are several qualitative features, however, that we want to emphasize and undoubtedly would remain if a full bulk 
calculation were possible,

First, it is clear from our calculations that Eq.~(\ref{pseudo}) overestimates the impact of retardation. 
This point was already made in Refs.~[\onlinecite{bauer12},\onlinecite{bauer13}]
through a combination of nonperturbative Dynamical Mean Field Theory (DMFT) and perturbative calculations. The present calculations
suggest that {\it no pairing occurs for large values of the Coulomb repulsion}, in contrast to what Eq.~(\ref{pseudo}) would predict.
In Ref.~[\onlinecite{bonca00}] a strong coupling argument was made to show that binding was limited to 
$U< 8\lambda t$. We find that
in practice binding ceases at values of $U$ even lower than this estimate, as the examples in Fig.~\ref{fig4_2site} illustrate.

A second feature is the non-monotonicity of the binding as a function of $\lambda$ (with $U \propto \lambda t$). Again this is in contrast to what
Eq.~(\ref{pseudo}) would predict, where the binding would continue to increase as $\lambda $ is increased, even if $U \propto \lambda t$, since the pseudopotential
argument would eventually render the large $U$ to be relatively harmless. Also, unlike standard Eliashberg calculations, the binding here is
zero for sufficiently small values of $\lambda$, as shown in Fig.~\ref{fig5_2site}. 

A third feature is the polaronic nature of the single-particle ground state. Unless such characteristics are present in the system of interest (for most
known superconductors they are not), then the parameter regime is further restricted in this regard. This is evident here because we have been
able to access more realistic low phonon frequencies, where such polaronic tendencies are enhanced.

On the other hand, with such a small system, we are unable to make an assessment of the competition for antiferromagnetic and charge-density-wave 
order. Other researchers have weighed in with regards to this competition.\cite{nowadnick12,karakuzu17,bradley21} We note that the primary issue investigated
here, retardation, appears to favor antiferromagnetic correlations (see for example, Fig.~2 of Ref.~[\onlinecite{nowadnick12}]). As mentioned in the
introductory discussion, however, the quantum Monte Carlo method used there made it difficult to explore the $\omega_E << t$ regime. We are also unable to
say anything about the Migdal approximation, since we obviously cannot assemble a Fermi sea to achieve the desired goal of $\hbar \omega_E <<E_F$,
where $E_F$ is the Fermi energy.

\section{Summary}

The main message of this paper, already noted to some extent in previous work,\cite{marsiglio95,bonca00,berciu07,bauer12,bauer13} is that
the Bogoliubov-Morel-Anderson pseudopotential renormalization suggested by Eq.~(\ref{pseudo}) is not very accurate. In particular this equation
does not contain the notion of a maximum value of Coulomb repulsion, beyond which no pairing occurs. The two-site calculations presented
here highlight this deficiency.

Beyond this message our calculations, while exact, can only be suggestive of what will actually occur in the bulk limit in higher dimensions, and with
a macroscopic number or particles. We hope to provide further progress on some of these issues in future work.

\begin{acknowledgments}
I am grateful to Jorge Hirsch for initial discussions and calculations that prompted this study, and for subsequent constructive criticism.
This work was supported in part by the Natural Sciences and Engineering
Research Council of Canada (NSERC) and by an MIF from the Province of Alberta.

\end{acknowledgments}

\vskip0.2in

\appendix

\section{The two-site Hamiltonian and Hilbert space}

We can now gather up the various contributions to Eq.~(\ref{hh_ham1}) and write them down for the two-site model:
\begin{eqnarray}
H = &-& 2t \left[c^\dagger_{1\uparrow} c_{2\uparrow}^{\phantom{\dagger}} + c^\dagger_{2\uparrow} c_{1\uparrow}^{\phantom{\dagger}} + c^\dagger_{1\downarrow} c_{2\downarrow}^{\phantom{\dagger}} + c^\dagger_{2\downarrow} c_{1\downarrow}^{\phantom{\dagger}} \right]
\nonumber \\
 &+& \hbar \omega_E \left[ a_1^\dagger a_1^{\phantom{\dagger}} + a_2^\dagger a_2^{\phantom{\dagger}} \right]
\nonumber \\
 &-& g \hbar \omega_E \left[ (n_{1\uparrow} + n_{1\downarrow}) (a_1^\dagger + a_1^{\phantom{\dagger}}) + (n_{2\uparrow} + n_{2\downarrow}) 
 (a_2^\dagger + a_2^{\phantom{\dagger}})\right]
\nonumber \\
 &+& U  \left[ n_{1\uparrow} n_{1\downarrow} + n_{2\uparrow} n_{2\downarrow} \right],
\label{hh_ham2}
\end{eqnarray}
where the $2t$ occurs in the first line (rather than merely $t$) because of the periodic boundary conditions.
One can straightforwardly solve this problem with a Hilbert space consisting of enumerated electron states, with fixed electron number $N_e$
($N_e = 0$, 1, 2, 3 or 4), in a direct product with the phonon Fock states on site 1, $N_{ph 1} = 0,1,2,...N_{\rm max}$ and on site 2, 
$N_{ph 2} = 0,1,2,...N_{\rm max}$, where $N_{\rm max}$ represents a truncation at $N_{\rm max} + 1$ phonon Fock states at each site. This
number is to be increased until the results of interest are converged.

A more efficient procedure\cite{ranninger92} is to define operators
\begin{equation}
d_0 \equiv {a_1 + a_2 \over \sqrt{2}}, \ \ \ \ d_\pi \equiv {a_1 - a_2 \over \sqrt{2}},
\label{s_and_d}
\end{equation}
and similarly for $d_0^\dagger$ and $d_\pi^\dagger$. For the electron operators, we define
\begin{equation}
c_{0\sigma} \equiv {c_{1\sigma} + c_{2\sigma} \over \sqrt{2}}, \ \ \ \ c_{\pi \sigma} \equiv {c_{1\sigma} - c_{2\sigma} \over \sqrt{2}},
\label{c0_and_cpi}
\end{equation}
and similarly for $c^\dagger_{0\sigma}$ and $c^\dagger_{\pi \sigma}$.
Then, straightforward algebra yields
\begin{widetext}
\begin{eqnarray}
H = &-& 2t \left[c^\dagger_{0\uparrow} c_{0\uparrow}^{\phantom{\dagger}} + c^\dagger_{0\downarrow} c_{0\downarrow}^{\phantom{\dagger}} - 
c^\dagger_{\pi \uparrow} c_{\pi \uparrow}^{\phantom{\dagger}} - c^\dagger_{\pi \downarrow} c_{\pi \downarrow}^{\phantom{\dagger}} \right]
\nonumber \\
 &+& \hbar \omega_E \left[ d_{0}^\dagger d_0^{\phantom{\dagger}} + d_\pi^\dagger d_\pi^{\phantom{\dagger}} \right]
\nonumber \\
 &-& {g \hbar \omega_E \over \sqrt{2}}
 \left[ N_{e}(d_0^\dagger + d_0^{\phantom{\dagger}})
 + (c^\dagger_{0\uparrow} c_{\pi \uparrow}^{\phantom{\dagger}} + c^\dagger_{\pi \uparrow} c_{0\uparrow}^{\phantom{\dagger}}
 + c^\dagger_{0\downarrow} c_{\pi \downarrow}^{\phantom{\dagger}} + c^\dagger_{\pi \downarrow} c_{0\downarrow}^{\phantom{\dagger}}) 
 (d_{\pi}^\dagger + d_\pi^{\phantom{\dagger}})\right]
\nonumber \\
 &+& {U\over 2}  \left[ n_{0\uparrow} n_{0\downarrow} + n_{\pi \uparrow} n_{\pi \downarrow}  + n_{0 \uparrow} n_{\pi \downarrow} 
 + n_{\pi \uparrow} n_{0 \downarrow} + p_{0\pi\uparrow} p_{0\pi\downarrow}   + p_{\pi 0\uparrow} p_{\pi 0\downarrow}
  + p_{0\pi\uparrow} p_{\pi 0\downarrow} + p_{\pi 0\uparrow} p_{0\pi\downarrow}  \right],
\label{hh_ham3}
\end{eqnarray}
where $p_{k k^\prime\sigma}^{\phantom{\dagger}} \equiv c_{k\sigma}^\dagger c_{k^\prime\sigma}^{\phantom{\dagger}}$ represents a
wave vector transfer from $k^\prime$ to $k$ (we set the lattice spacing $\ell = 1$). In the third line, $N_e$ is a constant, so this term can be combined with the first
term of the second line via
\begin{equation}
\tilde{d}_0 \equiv d_0 - {gN_e \over \sqrt{2}} \ \ \ {\rm and} \ \ \  \tilde{d}_0^\dagger \equiv d_0^\dagger - {gN_e \over \sqrt{2}} 
\label{shifted}
\end{equation}
so now the Hamiltonian becomes
\begin{eqnarray}
H = &-&\hbar \omega_E {g^2 N_e^2 \over 2} \nonumber \\
&-& 2t \left[c^\dagger_{0\uparrow} c_{0\uparrow}^{\phantom{\dagger}} + c^\dagger_{0\downarrow} c_{0\downarrow}^{\phantom{\dagger}} - 
c^\dagger_{\pi \uparrow} c_{\pi \uparrow}^{\phantom{\dagger}} - c^\dagger_{\pi \downarrow} c_{\pi \downarrow}^{\phantom{\dagger}} \right]
\nonumber \\
 &+& \hbar \omega_E \left[ \tilde{d}_{0}^\dagger \tilde{d}_0^{\phantom{\dagger}} + d_\pi^\dagger d_\pi^{\phantom{\dagger}} \right]
\nonumber \\
 &-& {g \hbar \omega_E \over \sqrt{2}}
% \left[
 (c^\dagger_{0\uparrow} c_{\pi \uparrow}^{\phantom{\dagger}} + c^\dagger_{\pi \uparrow} c_{0\uparrow}^{\phantom{\dagger}}
 + c^\dagger_{0\downarrow} c_{\pi \downarrow}^{\phantom{\dagger}} + c^\dagger_{\pi \downarrow} c_{0\downarrow}^{\phantom{\dagger}}) 
 (d_{\pi}^\dagger + d_\pi^{\phantom{\dagger}})
 %\right]
\nonumber \\
 &+& {U\over 2}  \left[ n_{0\uparrow} n_{0\downarrow} + n_{\pi \uparrow} n_{\pi \downarrow}  + n_{0 \uparrow} n_{\pi \downarrow} 
 + n_{\pi \uparrow} n_{0 \downarrow} + p_{0\pi\uparrow} p_{0\pi\downarrow}   + p_{\pi 0\uparrow} p_{\pi 0\downarrow}
  + p_{0\pi\uparrow} p_{\pi 0\downarrow} + p_{\pi 0\uparrow} p_{0\pi\downarrow}  \right],
\label{hh_ham4}
\end{eqnarray}
%\end{widetext}
and only the antisymmetric phonon degree of freedom ($d_\pi$) needs to be treated numerically, and a constant (binding) energy results
from the coupling of the electron and symmetric mode degrees of freedom. Moreover, this Hamiltonian is parity-conserving,
and is therefore block-diagonal in total parity. Each phonon degree of freedom ($d_\pi$) carries wave vector $\pi$, so the two sets of Hilbert space
have total wave vector $0$ or $\pi$, respectively. 
\end{widetext}

\subsubsection{One electron}

In the one electron spin-up sector, for example, the basis states are enumerated as
\begin{eqnarray}
q_{\rm tot} = 0  \quad & & \quad \quad \quad \quad \quad  q_{\rm tot} = \pi \nonumber \\
|0\rangle_0 \equiv c_{0\uparrow}^\dagger|0\rangle & & \quad \quad \quad \quad  |0\rangle_\pi \equiv c_{\pi\uparrow}^\dagger|0\rangle \nonumber \\
|1\rangle_0 \equiv c_{\pi\uparrow}^\dagger|1\rangle & & \quad \quad \quad \quad  |1\rangle_\pi \equiv c_{0\uparrow}^\dagger|1\rangle \nonumber \\
|2\rangle_0 \equiv c_{0\uparrow}^\dagger|2\rangle & & \quad \quad \quad \quad  |2\rangle_\pi \equiv c_{\pi\uparrow}^\dagger|2\rangle \nonumber \\
|3\rangle_0 \equiv c_{\pi\uparrow}^\dagger|3\rangle & & \quad \quad \quad \quad  |3\rangle_\pi \equiv c_{0\uparrow}^\dagger|3\rangle \nonumber \\
\vdots \quad \quad \quad && \quad \quad \quad \quad \quad \quad \ \vdots  \nonumber \\
|N_{\rm max}\rangle_0 \equiv c_{0\uparrow}^\dagger|N_{\rm max}\rangle & & \quad \quad \quad \quad  |N_{\rm max}\rangle_\pi \equiv c_{\pi\uparrow}^\dagger|N_{\rm max}\rangle \nonumber \\
%\vdots \quad \quad \quad && \quad \quad \quad \quad \quad \quad \ \vdots
\label{basis_states}
\end{eqnarray}
where a total of $N_{\rm max}+1$ states are used for each sector, one with $q_{\rm tot} = 0$ and one with $q_{\rm tot} = \pi$. For the case listed
$N_{\rm max}$ is even, and in each sector the ket $|n\rangle$ (without a subscript) enumerates the number of phonons. The normalized set has
\begin{equation}
|n\rangle \equiv {1 \over \sqrt{n!}} (d_\pi^\dagger)^n|0\rangle, \ \ \ n = 0, 1, 2, \ldots
\label{fock}
\end{equation}
and $|0\rangle$ is the phonon vacuum, while each set overall is enumerated by the kets with subscript $0$ and $\pi$, respectively.

Then we simply expand the one electron wave function in terms of this basis (say, for $q_{\rm tot} = 0$),
\begin{equation}
|\psi_{1e,0}\rangle = \sum_{n=0}^{N_{\rm max}} b_{n0}|n\rangle_0^{\phantom{a}}
\label{psi_1e_0}
\end{equation}
and the Schr\"odinger equation becomes the eigenvalue problem,
\begin{equation}
\sum_{m=1}^{N_{\rm max}} H_{nm} b_{m0} = Eb_{n0}
\label{schro_1e}
\end{equation}
where the matrix $H_{nm}$ is given simply as the tri-diagonal form,
\begin{eqnarray}
\begin{pmatrix}
    \epsilon_0(0) &  v(0)            & 0                & 0               &  0              & 0          & 0        &\dots \\
    v(0)          &  \epsilon_0(1)   & v(1)            & 0               &  0              & 0          & 0        &\dots  \\
     0              &  v(1)           &  \epsilon_0(2) & v(2)           & 0               & 0          & 0         & \dots  \\
     0              &  0               &  v(2)           &  \epsilon_0(3)  & v(3)           & 0         & 0        & \dots  \\
     0              &  0               &  0               &  v(3)           &  \epsilon_0(4) & v(4)     & 0        & \dots  \\
     \vdots       & \vdots        & \vdots        & \vdots         & \vdots        &\vdots  & \vdots & \ddots
%         x_{11} & x_{12} & x_{13} & \dots  & x_{1n} \\
%    x_{21} & x_{22} & x_{23} & \dots  & x_{2n} \\
%    \vdots & \vdots & \vdots & \ddots & \vdots \\
%    x_{d1} & x_{d2} & x_{d3} & \dots  & x_{dn}
   \end{pmatrix}
\label{ham_matrix_1e}
\end{eqnarray}
where 
\begin{eqnarray}
\epsilon_0(n) &=& - {g^2\hbar \omega_E \over 2}  + (-1)^{n+1}2t + n\hbar \omega_E \nonumber \\
v(n) &=& -{g\hbar \omega_E \over \sqrt{2}} \sqrt{n+1}.
\label{ham_defns_1e}
\end{eqnarray}
For $q_{\rm tot} = \pi$, one simply replaces Eq.~(\ref{psi_1e_0}) with
\begin{equation}
|\psi_{1e,\pi}\rangle = \sum_{n=0}^{N_{\rm max}} b_{n\pi}|n\rangle_\pi^{\phantom{a}}
\label{psi_1e_pi}
\end{equation}
and replaces $\epsilon_0(n)$ with $\epsilon_\pi(n)$ in Eq.~(\ref{ham_matrix_1e}) where
\begin{equation}
\epsilon_\pi(n) = - {g^2\hbar \omega_E \over 2}  + (-1)^{n}2t + n\hbar \omega_E.
\label{ham_defns_1e_pi}
\end{equation}

\subsubsection{Two electrons}

The two-electron states with total $S_z = 0$ are enumerated similarly to those of the one electron spin-up sector. They are
\begin{eqnarray}
q_{\rm tot} = 0  \quad \quad & & \quad \quad \quad \quad \quad  q_{\rm tot} = \pi \nonumber \\
%& & \nonumber \\
|1\rangle_0 \equiv c_{0\uparrow}^\dagger c_{0\downarrow}^\dagger   |0\rangle & & \quad \quad \quad \quad  |1\rangle_\pi \equiv c_{0\uparrow}^\dagger c_{\pi\downarrow}^\dagger |0\rangle \nonumber \\
|2\rangle_0 \equiv c_{\pi\uparrow}^\dagger c_{\pi\downarrow}^\dagger   |0\rangle & & \quad \quad \quad \quad  |2\rangle_\pi \equiv c_{\pi\uparrow}^\dagger c_{0\downarrow}^\dagger |0\rangle \nonumber \\
%& & \nonumber \\
|3\rangle_0 \equiv c_{0\uparrow}^\dagger c_{\pi\downarrow}^\dagger   |1\rangle & & \quad \quad \quad \quad  |3\rangle_\pi \equiv c_{0\uparrow}^\dagger c_{0\downarrow}^\dagger |1\rangle \nonumber \\
|4\rangle_0 \equiv c_{\pi\uparrow}^\dagger c_{0\downarrow}^\dagger   |1\rangle & & \quad \quad \quad \quad  |4\rangle_\pi \equiv c_{\pi\uparrow}^\dagger c_{\pi\downarrow}^\dagger |1\rangle \nonumber \\
%& & \nonumber \\
|5\rangle_0 \equiv c_{0\uparrow}^\dagger c_{0\downarrow}^\dagger   |2\rangle & & \quad \quad \quad \quad  |5\rangle_\pi \equiv c_{0\uparrow}^\dagger c_{\pi\downarrow}^\dagger |2\rangle \nonumber \\
|6\rangle_0 \equiv c_{\pi\uparrow}^\dagger c_{\pi\downarrow}^\dagger   |2\rangle & & \quad \quad \quad \quad  |6\rangle_\pi \equiv c_{\pi\uparrow}^\dagger c_{0\downarrow}^\dagger |2\rangle \nonumber \\
%& & \nonumber \\
|7\rangle_0 \equiv c_{0\uparrow}^\dagger c_{\pi\downarrow}^\dagger   |3\rangle & & \quad \quad \quad \quad  |7\rangle_\pi \equiv c_{0\uparrow}^\dagger c_{0\downarrow}^\dagger |3\rangle \nonumber \\
|8\rangle_0 \equiv c_{\pi\uparrow}^\dagger c_{0\downarrow}^\dagger   |3\rangle & & \quad \quad \quad \quad  |8\rangle_\pi \equiv c_{\pi\uparrow}^\dagger c_{\pi\downarrow}^\dagger |3\rangle \nonumber \\
%& & \nonumber \\
%|1\rangle_0 \equiv c_{\pi\uparrow}^\dagger|1\rangle & & \quad \quad \quad \quad  |1\rangle_\pi \equiv c_{0\uparrow}^\dagger|1\rangle \nonumber \\
%|2\rangle_0 \equiv c_{0\uparrow}^\dagger|2\rangle & & \quad \quad \quad \quad  |2\rangle_\pi \equiv c_{\pi\uparrow}^\dagger|2\rangle \nonumber \\
%|3\rangle_0 \equiv c_{\pi\uparrow}^\dagger|3\rangle & & \quad \quad \quad \quad  |3\rangle_\pi \equiv c_{0\uparrow}^\dagger|3\rangle \nonumber \\
\vdots \quad \quad \quad && \quad \quad \quad \quad \quad \quad \ \vdots  \nonumber \\
|N_{2m}-1\rangle_0 \equiv c_{0\uparrow}^\dagger c_{0\downarrow}^\dagger   |N_{\rm max}\rangle & & \quad  |N_{2m}-1\rangle_\pi \equiv c_{0\uparrow}^\dagger c_{\pi\downarrow}^\dagger |N_{\rm max}\rangle \nonumber \\
|N_{2m}\rangle_0 \equiv c_{\pi\uparrow}^\dagger c_{\pi\downarrow}^\dagger   |N_{\rm max}\rangle & & \quad \quad  \ \   |N_{2m}\rangle_\pi \equiv c_{\pi\uparrow}^\dagger c_{0\downarrow}^\dagger |N_{\rm max}\rangle \nonumber \\
%|N_{\rm max}\rangle_0 \equiv c_{0\uparrow}^\dagger|N_{\rm max}\rangle & & \quad \quad \quad \quad  |N_{\rm max}\rangle_\pi \equiv c_{\pi\uparrow}^\dagger|N_{\rm max}\rangle \nonumber \\
%\vdots \quad \quad \quad && \quad \quad \quad \quad \quad \quad \ \vdots
\label{basis_states2}
\end{eqnarray}
where now each sector has a total of $N_{2m} = 2N_{\rm max}+2$ states (here $N_{\rm max}$ is assumed to be even), and again the ket $|n\rangle$
without a subscript denotes the normalized phonon state given in Eq.~(\ref{fock}). The kets with a subscript simply enumerate the states; we have used
the convention that they begin at unity, whereas the single particle basis states given in Eq.~(\ref{basis_states}) used the convention that they begin at zero.
Now for the $q_{\rm tot} = 0$ subspace we use an expansion with coefficients $f_{no}$
\begin{equation}
|\psi_{2e,0}\rangle = \sum_{n=1}^{N_{2m}} f_{n0}|n\rangle_0^{\phantom{a}}
\label{psi_2e_0}
\end{equation}
and the two-particle Schr\"odinger equation becomes the eigenvalue problem,
\begin{equation}
\sum_{m=1}^{N_{2m}} H_{2[n,m]} f_{m0} = Ef_{n0}
\label{schro_2e}
\end{equation}
where the matrix elements are either diagonal or involve one phonon creation or annihilation. In addition the Hubbard $U$ interaction is
off-diagonal in this basis between states with the same number of phonons.
For example, states of the type $c_{0\uparrow}^\dagger c_{0\downarrow}^\dagger   |n\rangle$ have diagonal matrix elements
$H_{2[4n+1,4n+1]} = -4t + n\hbar \omega_E + U/2$. States of the type $c_{\pi\uparrow}^\dagger c_{\pi\downarrow}^\dagger   |n\rangle$ have
diagonal matrix elements $H_{2[4n+2,4n+2]} = +4t + n\hbar \omega_E + U/2$, while either $c_{0\uparrow}^\dagger c_{\pi\downarrow}^\dagger   |n\rangle$
or $c_{\pi\uparrow}^\dagger c_{0\downarrow}^\dagger   |n\rangle$ have the same diagonal matrix elements
$H_{2[4n+3,4n+3]} = H_{2[4n+4,4n+4]} = n\hbar \omega_E + U/2$. 

For the Hubbard interaction, $H_{2[4n+1,4n+2]} = H_{2[4n+2,4n+1]} =U/2$, and similarly $H_{2[4n+3,4n+4]} = H_{2[4n+4,4n+3]} =U/2$.
Finally, states differing by one phonon have non-zero matrix elements, and these are given by the usual square-roots generated by the phonon
creation and annihilation operators. For a given maximum number of phonons the two-electron Hamiltonian matrix has double the dimension of the
single particle matrix. As in the one-particle sector, an identical procedure applies to the $q_{\rm tot} = \pi$ subspace.

%\appendix
%
%\section{probably not needed}

\end{document}